\documentclass[12pt]{article}
\usepackage{stmaryrd}
\usepackage{amssymb}
\usepackage{graphics}
\usepackage{epsfig}
\usepackage{a4wide}
\usepackage{cite}
\usepackage{color}

\begin{document}

\title{Heating of ions by low-frequency Alfv\'{e}n waves in partially ionized plasmas}
\author{Chuanfei Dong$^\mathrm{a}$\thanks{%
dcfy@umich.edu} and Carol S. Paty$^\mathrm{b}$\thanks{%
carol.paty@eas.gatech.edu} \\
{$^\mathrm{a}${\small Department of Atmospheric, Oceanic and Space
Sciences, University of
Michigan,}} \\
{\small Ann Arbor, MI 48109, U.S.A.} \\
{$^\mathrm{b}${\small School of Earth and Atmospheric Sciences,
Georgia Institute of Technology,}} \\
{\small Atlanta, GA 30332, U.S.A.}}

\date{}
\maketitle \vskip 15mm
\begin{abstract}
In the solar atmosphere, the chromospheric and coronal plasmas are
much hotter than the visible photosphere. The heating of the solar
atmosphere, including the partially ionized chromosphere and corona,
remains largely unknown. In this paper we demonstrate that the ions
can be substantially heated by Alfv\'{e}n waves with very low
frequencies in partially ionized low beta plasmas. This differs from
other Alfv\'{e}n wave related heating mechanisms such as ion-neutral
collisional damping of Alfv\'{e}n waves and heating described by
previous work on resonant Alfv\'{e}n wave heating. In this paper, we
find that the non-resonant Alfv\'{e}n wave heating is less efficient
in partially ionized plasmas than when there are no ion-neutral
collisions, and the heating efficiency depends on the ratio of the
ion-neutral collision frequency to the ion gyrofrequency.

\end{abstract}



\maketitle

\renewcommand{\theequation}{\arabic{section}.\arabic{equation}}
\renewcommand{\thesection}{\Roman{section}.}
\newcommand{\nb}{\nonumber}

\vfill \eject

\baselineskip=0.32in

\section{INTRODUCTION}
\par
The chromosphere, the region between the solar surface and the
corona, is permeated by low-frequency (0.0014 Hz $\sim$ 0.0079 Hz)
Alfv\'{e}n waves with strong amplitudes (2.6 kilometers per second
in amplitude)\cite{sciDB} while the plasma beta
value($\beta=(v_p/v_A)^2$, where $v_p$ and $v_A$ are the thermal
speed and the Alfv\'{e}n speed, respectively) in this region is low
(where $\beta$ reaches a minimum with $\beta <
3\times10^{-2}$)\cite{sciBDe,solarG}. The chromospheric and coronal
plasmas are much hotter than the visible photosphere. The heating
mechanisms in these regions, however, have not yet been fully
understood\cite{sciDB,sciBDe,sciH}. Alfv\'{e}n waves have long been
considered to play a crucial role in heating of plasma in these two
regions and in magnetic fusion devices\cite{s1,s2}. Numerous
theoretical and experimental papers have been published to
investigate resonant heating of ions by Alfv\'{e}n
waves\cite{s4,res1,res2,res3,res4}. In these works, the cyclotron
resonant condition is necessary for ion heating by the Alfv\'{e}n
waves, and in general the frequencies of the applied Alfv\'{e}n
waves are comparable to the cyclotron frequency. However, the
heating of ions by low-frequency Alfv\'{e}n waves and related
problems have triggered great interest in recent years
\cite{YAPL,YDusty,lipop,wuPOP,chenPOP,wangPRL,wuPRL,luPOP,liAPL,SAPL,jaaprl0809}.
In these works, the ions can be heated by Alfv\'{e}n waves via
nonresonant interactions, which means $\omega \ll {\Omega}_{q}$,
where $\omega$ is the frequency of the Alfv\'{e}n waves and
${\Omega}_{q}$ denotes the gyrofrequency for ion species labeled
\emph{q}. Researchers have used various simulation and theoretical
methods in order to validate the heating mechanism of low-frequency
Alfv\'{e}n wave interactions; these include test particle approaches
\cite{wuPOP,wangPRL,luPOP,liAPL,YAPL}, two-fluid magnetohydrodynamic
(MHD) model (nonlinear parametric interaction of Alfv\'{e}n waves
with slow/fast waves)\cite{YDusty}, hybrid
simulations\cite{lipop,jaaprl0809}, and kinetic
theory\cite{wuPRL,jaaprl0809}. Importantly, several of these works
demonstrate quantitatively similar results even while employing
different approaches. For example, the numerical and analytical
results adopting the heating mechanism proposed by Wang and Wu
(discussed later in this paper)\cite{wangPRL,wuPRL} are
quantitatively in accordance with the work of Voitenko and Goossens
who examine cross-field heating by low-frequency kinetic Alfven
waves (KAWs)\cite{YAPL}.

\par
Some recent works\cite{DeAA98,DeAPJ01,LeakeAA} indicate that
Alfv\'{e}n waves can propagate through the partially ionized solar
chromosphere. They find that Alfv\'{e}n waves with frequencies below
0.6 Hz will not be completely damped. More importantly, these works
all show that Alfv\'{e}n waves with frequencies below 0.01 Hz are
undamped by ion-neutral collisions in the solar chromosphere, and
are therefore available for damping by other mechanisms, such as the
mechanism described in this paper.

\par
In this paper, we demonstrate that the low-frequency Alfv\'{e}n
waves propagating along the background magnetic field
${\textbf{B}}_{0}={B}_{0}{\textbf{i}}_{z}$, can heat ions even in
partially ionized plasmas. It is important to note the heating
mechanism in this paper is different from the previous
works\cite{LeakeAA,DeAA98,DeAPJ01,RS}. The heating process in the
present work is due to a randomization of the spatial velocity
distribution in the ion population, which is caused by the
nonresonant Alfv\'{e}n wave interactions. We find that the heating
process becomes less efficient than the situation with no
ion-neutral collisions. Moreover, the nonresonant heating process is
only effective for low-beta plasmas\cite{wangPRL} and Alfv\'{e}n
wave frequencies lower than 0.6 Hz; the most efficient heating
occurs when $\omega \leq 0.01$ Hz since low frequency Alfv\'{e}n
waves are more abundant due to the damping of high frequency waves
by ion-neutral collisions\cite{LeakeAA,DeAA98,DeAPJ01}.

\vskip 10mm

\section{ANALYTIC THEORY AND TEST PARTICLE CALCULATIONS}
\par
We consider the Alfv\'{e}n waves have a spectrum and the dispersion
relation can be described as $\omega=k{v}_{A}$ (${v}_{A}$ is the
Alfv\'{e}n speed, $\omega$ and $k$ are the wave angular frequency
and wave number, respectively). This relationship is still
appropriate even when the plasma is partially ionized, as described
in the following paragraphs. Without loss of generality, we consider
left-hand circular polarization in this paper. The wave magnetic
field vector $\delta{\textbf{B}}_{w}$ and electric field vector
$\delta{\textbf{E}}_{w}$ can be expressed as
\begin{equation}
\delta{\textbf{B}}_{w}=\sum_{k}{B}_{k}(\cos{\phi}_{k}{\textbf{i}}_{x}-\sin{\phi}_{k}{\textbf{i}}_{y}),
\end{equation}
\begin{equation}
\delta{\textbf{E}}_{w}=-\frac{v_A}{c} \textbf{b} \times
\delta{\textbf{B}}_{w}, ~~~\textbf{b}=\frac{\textbf{B}_0}{B_0}
\end{equation}
where ${\textbf{i}}_{x}$ and ${\textbf{i}}_{y}$ are unit directional
vectors, ${\phi}_{k}=k({v}_{A}t-z)+\varphi_k$ denotes the wave phase
and $\varphi_k$ is the random phase for mode $k$. In the following,
we pay attention to the protons only, whose equation of motion is
described by
\begin{equation}
\label{vtot}
m_i\frac{d\textbf{v}}{dt}=q_i\left(\delta{\textbf{E}}_{w}+\frac{\textbf{v}}{c}\times({\textbf{B}}_{0}+\delta{\textbf{B}}_{w})\right)+m_i{\nu}_{in}(\textbf{u}-\textbf{v}),~~~\frac{d\textbf{r}}{dt}=\textbf{v}
\end{equation}
where $\textbf{v}$ is the ion velocity, $\textbf{u}$ is the bulk
velocity of a background neutral fluid, and ${\nu}_{in}$ is the
frequency for elastic collisions between ions and neutrals. Ion
collisions with electrons are neglected in Eq.(\ref{vtot}) because
$m_i \gg m_e$. The collision frequency responsible for momentum
transfer between species $i$ and $n$ is defined as:
\begin{equation}
\label{nuin}
{\nu}_{in}=\frac{m_n}{m_i+m_n}{n}_{n}\sqrt{\frac{8{k}_{B}T}{\pi{m}_{in}}}{\sigma}_{in},
\end{equation}
with ${m}_{in}=(m_im_n)/(m_i+m_n), {\sigma}_{in}$ the collisional
cross-section for collisions between the two species. Here $m_i=m_n$
because the plasma we studied in this paper is assumed to be
entirely composed of hydrogen which leads to ${m}_{in}=m_i/2$. The
wave phase speed ${v}_{ph}$ here is defined by\cite{DeAA98}:
\begin{equation}
\label{vph} v_{ph} = \frac{\omega}{k} =
v_A\sqrt{1-i\frac{\rho_n}{\rho_{tot}}\frac{\omega}{\nu_{ni}}}
\end{equation}
in which $\rho_n$ is the mass density of neutrals, $\rho_{tot}$ is
the total mass density of the plasma, and $\nu_{ni}$ is neutral-ion
collision frequency. Therefore, the relationship $\rho_n/\rho_{tot}
< 1$ is necessarily valid.  As shown by De Pontieu and Haerendel,
for waves with $\nu \leq 1$ Hz, the assumption $\omega \ll \nu_{ni}$
($\omega/\nu_{ni} \ll 1$) holds throughout the
chromosphere\cite{DeAA98}. Furthermore, most theories for the
generation of Alfv\'{e}n waves in the solar atmosphere predict
typical frequencies below 1 Hz\cite{DeAA98}, which has also been
observed\cite{sciDB}. These allow us to simplify Eq.(\ref{vph}) to
\begin{equation}
v_{ph} = \frac{\omega}{k} =
v_A\sqrt{1-i\frac{\rho_n}{\rho_{tot}}\frac{\omega}{\nu_{ni}}}
\approx v_{A}
\end{equation}
Hence the relationship $\omega=kv_A$ is still appropriate.

\par
Defining ${v}_{\perp}={v}_{x}+i{v}_{y}$,
${u}_{\perp}={u}_{x}+i{u}_{y}$, ${v}_{\parallel}={v}_{z}$,
${u}_{\parallel}={u}_{z}$ and $\delta
B_\omega=\sum_{k}B_ke^{-i\phi_k}$; we are left with
\begin{equation}
\label{vperp}\frac{dv_{\perp}}{dt}+(i\Omega_0+\nu_{in})v_{\perp}=i(v_\parallel-v_A)\sum_{k}\Omega_ke^{-i\phi_k}+\nu_{in}u_{\perp}
\end{equation}
\begin{eqnarray}
\label{vpara} \frac{dv_{\parallel}}{dt}
&=&-Im(v_\perp\sum_{k}\Omega_ke^{i\phi_k})+\nu_{in}(u_{\parallel}-v_\parallel),~~~\frac{dz}{dt}=v_\parallel
\end{eqnarray}
where $\Omega_0=\frac{q_iB_0}{m_ic}$ (the proton gyrofrequency),
$\Omega_k=\frac{q_iB_k}{m_ic}$. $Im(~)$ denotes the imaginary part
of its argument. As a first-order approximation, we can assume
$v_\parallel \approx v_\parallel(0)$, where $v_\parallel(0)$ is the
particle¡¯s initial parallel velocity. The approximation is valid
when $\frac{\Omega_k}{\Omega_0}=\frac{B_k}{B_0}$ is small enough and
the frequencies of the Alfv\'{e}n wave are sufficiently low to
ensure that $\mid\Omega_0\mid \gg \mid
k\left(v_\parallel(0)-v_A\right)\mid$. For simplicity, we assume the
bulk velocities of the cold neutrals $u_x(t)=u_y(t)=u_z(t) \approx
0$ and ion-neutral collision frequency $\nu_{in} \approx constant$,
which provides a lower limit for the amount of ion heating. Since
the ions and background neutral fluid are both ``cold'' initially,
it is reasonable to assume $\textbf{u}(t) \approx 0$ due to the
relatively short time scales (compared to the time required to heat
the neutrals) considered in this paper and the high neutral fraction
(detailed below).

\par
The ion gyrofrequency in the region of 800km $\sim$ 1500km above the
photosphere in solar chromosphere ranges from $1\times10^5 \sim
5\times10^5$Hz. The ion-neutral frequency in this region ranges from
$5\times10^3 \sim 5\times10^5$Hz. These results are based on the
solar atmospheric model VAL C\cite{APJS} for the densities and
temperatures, and a magnetic flux tube model with 1500G field
strength in the photosphere and 10G magnetic field strength in the
corona\cite{Bartphd}. We select two values of the ratio $a$
($a=\frac{\nu_{in}}{\Omega_0}$), 0.1 and 0.5, to represent the
region around 1000km above the photosphere. According to VAL C
model\cite{APJS} and the magnetic flux tube model\cite{Bartphd}
described above, when $a=\nu_{in}/\Omega_0 \geq 0.05$, the ratio of
neutral density to ion density $\lambda =n_n/n_i \gg 1$, which
indicates that the amount of neutrals is much larger than that of
ions. It is important to note that although the value of ratio
$\lambda \gg 1$ indicates that $\nu_{in}$ is large (with respect to
Alfv\'{e}n wave frequency $\omega$; refer to Eq.(\ref{nuin})), the
value of $\Omega_0$ can also be large (with respect to $\omega$) due
to the background magnetic field $B_0$
($\Omega_0=\frac{q_iB_0}{m_ic}$), thus the value of ratio $a$ can
still be very small.

\par
In the following, we focus on the perpendicular velocity component
due to the fact that the ion temperature increase is more prominent
along the perpendicular direction than the parallel
direction\cite{wangPRL,luPOP,YAPL}. We acknowledge that the
anisotropy between $v_\perp$ and $v_\parallel$ should be alleviated
after a sufficiently long period of time due to the collisional
effects, however, this is not taken into account due to the
relatively short time scales considered in this paper. Even over
much longer timescales, where ions temperatures become isotropic,
the fact that the ions are heated by Alfv\'{e}n waves remains
unchanged. With the initial condition $v_\perp=v_\perp(0)$ and
$z=z(0)$, the solution of Eq.(\ref{vperp}) can be written as:
\begin{eqnarray}
\label{insolution} v_\perp &=& v_{\perp}(0)e^{-(i\Omega_0+\nu_{in})t}+\frac{\nu_{in}u_{\perp}(\nu_{in}-i\Omega_0)}{{\Omega_0}^2+{\nu_{in}}^2}\left[1-e^{-i\Omega_0t}e^{-\nu_{in}}\right] \nb \\
&-&\frac{\sum_{k}\Omega_k\left[v_{A}-v_\parallel(0)\right](\Omega_0+i\nu_{in})}{{\Omega_0}^2+{\nu_{in}}^2}\left[e^{-ik(v_At-z)-i\varphi_k}-e^{i\left(kz(0)-\varphi_k\right)}e^{-(i\Omega_0+\nu_{in})t}\right]
\end{eqnarray}
Here, we use the approximation that
$\Omega_0-k[v_A-v_\parallel(0)]\approx\Omega_0$ and
$z=z(0)+v_\parallel(0)t$. In order to verify that the analytical
solution is correct, we set $\nu_{in}=0$. Then the Eqs. (\ref{vtot},
\ref{vperp}, \ref{vpara}) are reduced to those found in previous
work\cite{wangPRL,luPOP,wuPOP} where the analytical result is as
follows:
\begin{eqnarray}
\label{solution} v_\perp =
v_{\perp}(0)e^{-i\Omega_0t}-v_A\frac{\sum_{k}B_k}{B_0}e^{-ik(v_At-z)-i\varphi_k}+v_{A}\frac{\sum_{k}B_k}{B_0}e^{i\left(kz(0)-\varphi_k\right)}e^{-i\Omega_0t}
\end{eqnarray}
The relationships below between the kinetic temperature and velocity
of protons are based on plasma which consists of an ensemble of
protons,
\begin{eqnarray}
T_\perp(t)=\frac{1}{2}\sum_{i=1}^{N}
\frac{m_iv_{\perp}^{2}(t)}{N};~~~T_\parallel(t)=\sum_{i=1}^{N}\frac{m_iv_{\parallel}^{2}(t)}{N}
\end{eqnarray}
where $T_\perp$ is the perpendicular kinetic temperature,
$T_\parallel$ is the parallel kinetic temperature and $N$ is total
number of the protons.

\par
We present the simulation results using test-particle calculations
that build upon previous works\cite{wuPOP,wangPRL,luPOP,liAPL}. The
test-particle simulation will be valid in partially ionized plasmas
when no ion-neutral collisional damping occurs. We discretize the
Alfv\'{e}n wave number by
$k_j=k_{min}+(j-1)\frac{k_{max}-k_{min}}{J-1}$, for $j=1,...,J$,
where $k_{min}=k_1=1\times10^{-8}\Omega_0/v_A$ and
$k_{max}=k_J=5\times10^{-8}\Omega_0/v_A$. This range of wave numbers
implies that we are considering
$1\times10^{-8}\Omega_0<\omega<5\times10^{-8}\Omega_0$, so that the
wave frequencies are much lower than the proton gyrofrequency and
also will be low enough to guarantee that no ion-neutral collisional
damping occurs\cite{DeAA98,DeAPJ01,LeakeAA}. The field amplitudes of
different wave modes are equal to each other, and they are constant.
Here we set two values of $\delta B_{w}^{2}/B_{0}^{2}=\zeta \sim$
0.05 and 0.12. The total number of test particles is $10^5$, which
are randomly distributed during the time interval $0< \Omega_0 t <
2\pi$ and along the spatial range $0< z\Omega_0/v_A< 3\times10^9$.
In this paper, we vary the ratio $a$ ($a=\frac{\nu_{in}}{\Omega_0}$)
between 0.1 and 0.5, which is well within the acceptable range for
describing the solar chromosphere as determined by the previous
works\cite{APJS,Bartphd}. The initial ion velocities are assumed to
have a Maxwellian distribution with thermal speed $v_p$, which is
less than $v_A$ to ensure that the cyclotron resonance condition
cannot be satisfied.
\begin{figure}
\centering
\includegraphics[scale=0.5]{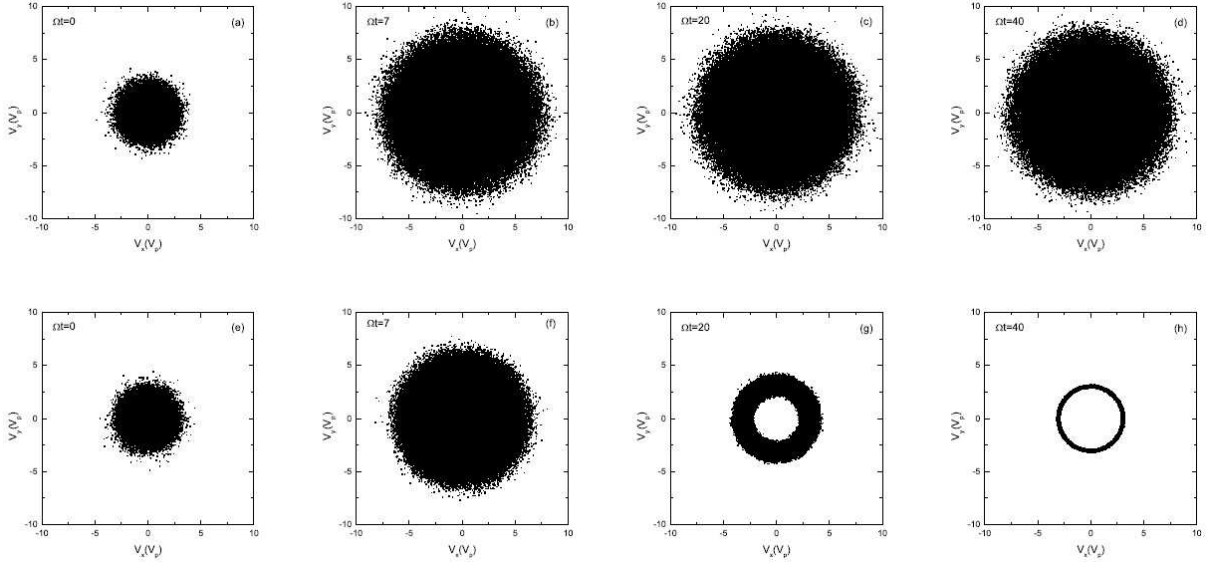}
\caption{\label{fig1} Velocity scatter plots of the test particles
in the $v_x-v_y$ space in the case $\nu_{in}=0$ (first line) and
$\nu_{in}=0.1\Omega_0$ (second line) at time $\Omega t$ = 0, $\Omega
t$ = 7, $\Omega t$ = 20 and $\Omega t$ = 40, for input parameters
($\delta{B_w}^{2}/{B_0}^{2}, v_p/v_A$) = (0.05, 0.07).}
\end{figure}

\par
In Fig.1 we present scatter plots in the $v_x-v_y$ space which
illustrate the process of particle heating. Here, the velocity is
normalized to the initial thermal speed $v_p$, which is set to be
$v_p$ = 0.07$v_A$ ($\beta=(v_p/v_A)^2=4.9\times10^{-3}\ll1$, which
is well within the range of chromosphere beta value determined by
Gary\cite{solarG}). We compare nonresonant heating without and with
neutral collisions in Fig.1. The results for the non-collisional
case, where $a=0$, shown in Fig.1 (a), (b), (c), and (d), are in
accordance with former results\cite{wangPRL}. We find that
ion-neutral collisions reduce the amount of heating over time and
are responsible for creating a ring distribution. Since
$\textbf{u}(t) \approx 0$ is valid during the relatively short time
scales considered in this paper as explained above, a balance is
reached between energy gained via Alfv\'{e}n wave interactions and
lost from collisions with cold neutrals. Thus the velocity
distribution evolves into a ``thin'' ring structure, with all of the
ions evolving to a single speed. This result can also be
analytically approximated, and will be discussed in detail in the
following paragraphs.
\begin{figure}
\centering
\includegraphics[scale=0.4]{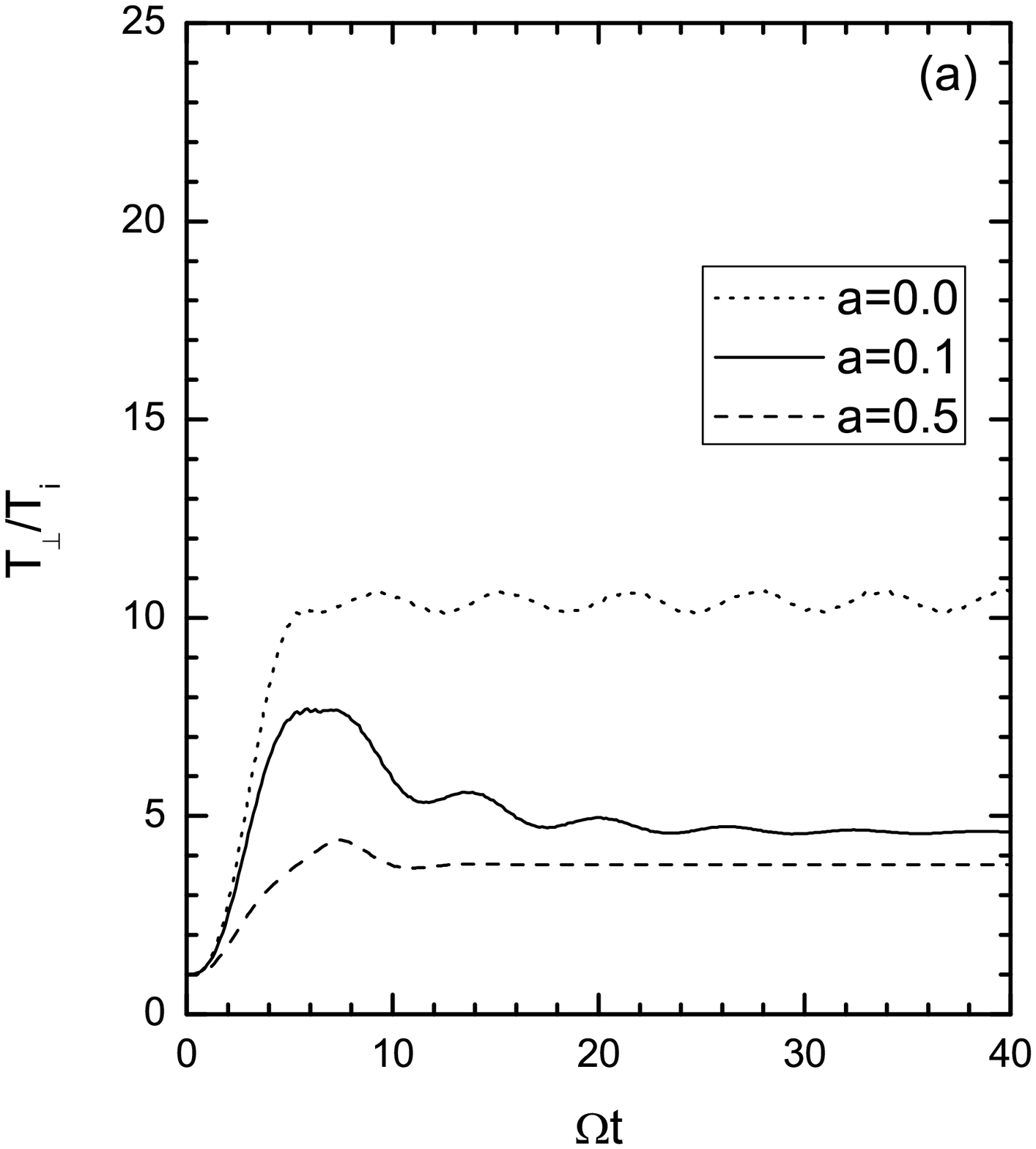}
\includegraphics[scale=0.4]{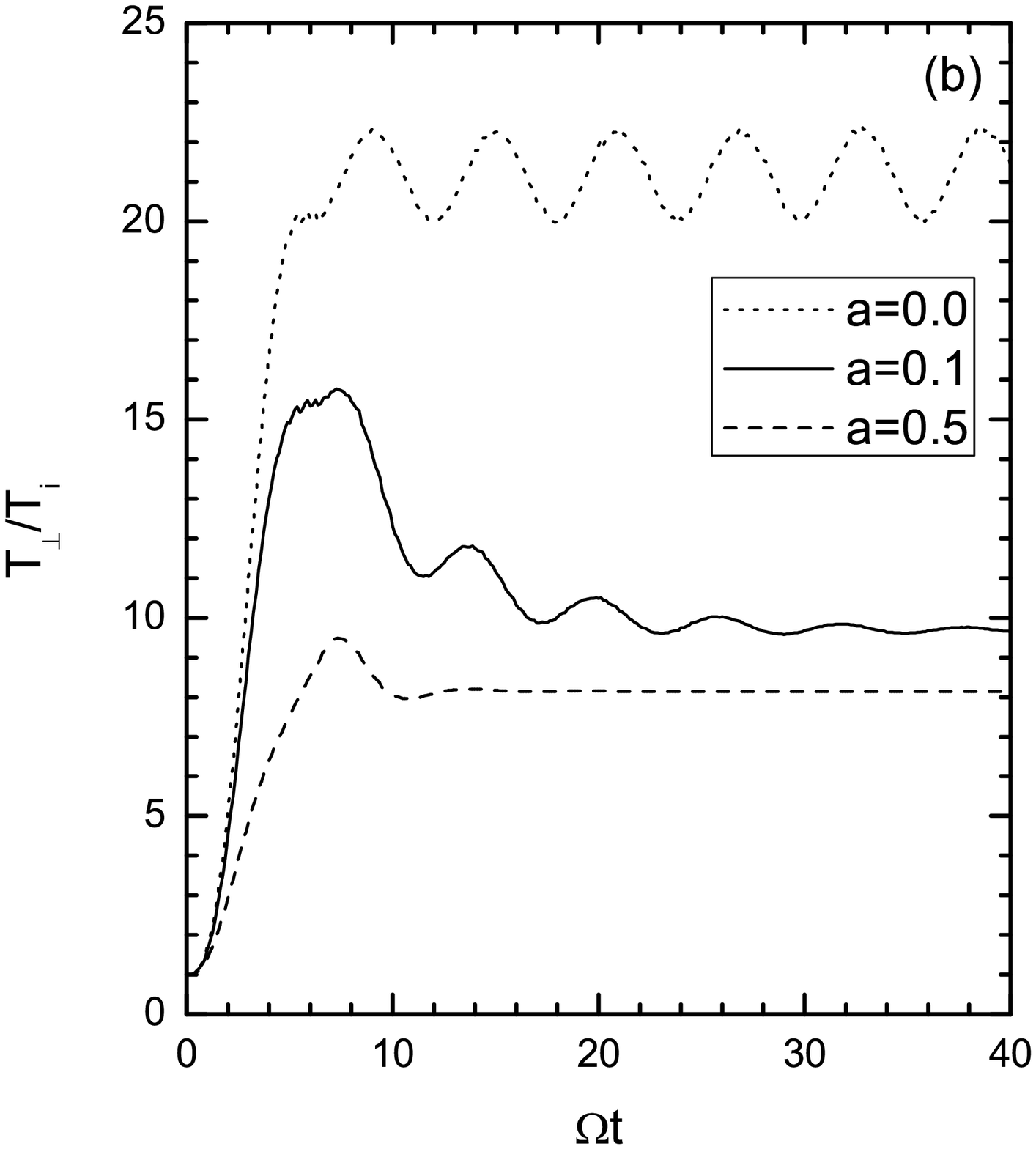}
\caption{\label{fig2} The temporal evolution of the perpendicular
kinetic temperatures normalized with respect to their initial values
$T_i$. The various input parameters, ($\delta{B_w}^{2}/{B_0}^{2},
v_p/v_A$)=(0.05, 0.07) and (0.12, 0.07) are shown in Fig.2(a) and
Fig.2(b), respectively.}
\end{figure}

\par
Fig.2 shows the temporal evolution of the kinetic temperatures,
where the results are based on two sets of input parameters
($\delta{B_w}^{2}/{B_0}^{2}, v_p/v_A$)=(0.05, 0.07) and (0.12,
0.07). The cases $a$ = 0, 0.1 and 0.5 are represented by the dotted
line, the solid line and the dashed line, respectively. The results
again demonstrate that heating via nonresonant Alfv\'{e}n wave
interactions is possible, and that even in the presence of neutrals
heating still occurs and reaches a steady value. Fig.2 illustrates
that the stronger Alfv\'{e}n wave amplitude
$\sqrt{\frac{\delta{B_w}^{2}}{{B_0}^{2}}}$ will result in the larger
amounts of heating, which is also indicated by Eq.(\ref{insolution},
\ref{solution}) and consistents with the previous
works\cite{wangPRL,wuPRL,luPOP}. We also find the ratio of the
ion-neutral collision frequency to the ion gyrofrequency, $a$, will
directly affect the heating process. When the ratio $a$ is larger,
the heating process becomes less efficient. However, the low-beta
plasmas are still significantly heated, as indicated in Fig.2 where
the temperature $T_\perp(t)$ is much larger than the initial
temperature $T_i$. This phenomenon may provide a partial explanation
as to why the temperature of the chromosphere is higher than the
photosphere while lower than the corona, as the fraction of ionized
particles increases from the photosphere through the chromosphere
and to the corona\cite{APJS}. Furthermore, we can see from the
evolution of the population in Fig.1(e), (f), (g), and (h) that
there exists a dynamic equilibrium between the heating and cooling
of the ions, which is reached after a period of time. This is also
demonstrated by the oscillations in temperature which decrease over
time as shown in Fig.2 for the cases $a\neq0$.

\par
The results of the test particle calculations are strongly
consistent with our analytical predictions. The ratio of frequencies
$a$ in the analytic solution Eq.(\ref{insolution}) is in the term
$e^{-\nu_{in}t}$ where $\nu_{in} = a\Omega_{0}$; therefore, a larger
value of $a$ will result in a lower value of $e^{-\nu_{in}t}$, which
will affect the value of ion kinetic temperatures as shown in Fig.2.
Given the parameters $u(t)\approx0$, $v_p=0.07v_A$, $\delta
B_{w}^{2}/B_{0}^{2}=\zeta=0.05$, $a=0.1$ and $t=40$, the solution of
Eq.(\ref{insolution}) can be reduced to
\label{vstable}\begin{eqnarray} \left|v_\perp \right| &\approx&
\left|
-\frac{\sum_{k}\Omega_k\left[v_{A}-v_\parallel(0)\right](\Omega_0+i\nu_{in})}{{\Omega_0}^2+{\nu_{in}}^2}e^{-ik(v_At-z)-i\varphi_k}
\right| \nb \\
&\approx& \left| \frac{\sqrt{\zeta} \left[v_{A}-v_\parallel(0)\right](1+ia)}{1+a^2}e^{-ik(v_At-z)-i\varphi_k} \right| \nb \\
&\approx& \sqrt{\zeta} \times v_{A}\left|
(1+ia)e^{-ik(v_At-z)-i\varphi_k} \right| \approx \sqrt{\zeta} \times
v_{A}\sqrt{1+a^2} \approx 3.2v_p
\end{eqnarray}
where we use the approximations $e^{-\varepsilon}\approx0$ when
$\varepsilon>3$, $|v_A-v_\parallel(0)|\approx |v_A|$ and
$1+a^2\approx1$. Thus the stable value of velocity is $3.2v_p$ which
is in accordance with the ring structure of velocity distribution
shown in Fig.1(h).

\par
The physical mechanism of this heating can be described as follows:
in low-beta plasma, nonresonant wave particle scattering by
Alfv\'{e}n waves can lead to randomization of the particle motion
transversely to the background magnetic field and thus effectively
to heating of the ions. Due to the fact that the neutrals are given
an approximately invariable bulk velocity, the heated ions will lose
energy from ion-neutral elastic collisions throughout the process.
This leads to an overall reduced efficiency of the ion heating
relative to the simulation with no ion-neutral collisions.

\vskip 10mm

\section{CONCLUSION}
\par
In summary, we show that partially ionized low beta plasmas can be
heated by a spectrum of low-frequency (0.0014 Hz $\sim$ 0.0079 Hz)
Alfv\'{e}n waves with large amplitude (2.6 kilometers per second in
amplitude), which have been observed in the low-beta (where $\beta$
reaches a minimum with $\beta < 3\times10^{-2}$\cite{solarG})
chromosphere\cite{sciDB}. This is contrary to the linear theory,
according to which ions can only be heated through resonant
interactions with Alfv\'{e}n waves. In our model, the frequencies of
the Alfv\'{e}n waves are much lower than the ion cyclotron
frequency, so the cyclotron resonant condition is not met. This also
ensures that the waves will not be damped by ion-neutral
collisions\cite{DeAA98,DeAPJ01,LeakeAA}. We find that the amount of
heating depends on the ratio $a$, where less heating occurs when the
importance of ion-neutral coupling is increased. However, in all
cases, significant heating of the ions occurred. Furthermore, we
show that the velocity distribution will form a ring structure
during the heating process when elastic ion-neutral collisions are
considered, which is consistent with our analytic approximation. Our
results demonstrate significant heating of ions through nonresonant
Alfv\'{e}n wave interactions in partially ionized plasmas, which may
provide an alternate source of heating in the solar chromosphere
than previously considered.

\vskip 10mm
\par
\noindent{\large\bf Acknowledgments:} The authors would like to
thank two anonymous referees, whose comments and suggestions greatly
improved the quality of this paper. C. F. Dong also appreciates many
fruitful discussions with Prof. C. B. Wang, Dr. C. S. Wu, Dr. B. De
Pontieu, Prof. T. D. Arber, Dr. M. L. Khodachenko, and Prof. Y. Li.
This work has been supported by the Georgia Institute of Technology.

\end{document}